# Simplistic approach for Determining Center Shift of Mössbauer Spectrum: exemplified on two cases

Stanisław M. Dubiel[1*] and Jan Żukrowski[2]

[1]Faculty of Physics and Applied Computer Science, [2]Academic Center for Materials and Nanotechnology, [1,2]AGH University of Science and Technology, PL-30-059 Kraków, Poland

## Abstract

A very simple method for determining the center (isomer) shift, *CS*, of a Mössbauer spectrum is outlined. Its applicability is demonstrated on two examples viz. pyrite and a ternary sigma-phase Fe-Cr-Ni compound. Sets of the spectra recorded in the temperature interval of 78-295 K for the former, and 5-293 K for the latter were analyzed with the simple and a commonly used methods. *CS(T)*-values obtained with both ways of the fitting procedures were analyzed in terms of the Debye model. The determined therefrom values of the Debye temperatures agree within the error limit with each other proving thereby that this very simple method gives correct values of *CS*.

\* Corresponding author: Stanislaw.Dubiel@fis.agh.edu.pl

## 1. Introduction

Among three possible spectral parameters characteristic of a Mössauer spectrum, a center shift, *CS*, also known as an isomer or chemical shift, is always present. The presence of the other two viz. the quadrupole splitting, *QS*, and the hyperfine magnetic field, *H*, depends on sample's properties. Namely, *QS* is present if the crystal symmetry of the sample is lower than cubic, and *H* occurs if the sample is magnetic or a non-magnetic sample is exposed to an external magnetic field. The scientific interest in the knowledge of *CS* is twofold: first, it gives information on the density of *s*-like electrons at a probe nucleus e. g. $^{57}$Fe, $^{119}$Sn, $^{151}$Eu, second, its temperature dependence, *CS(T)*, is related to lattice vibrations and permits determination of the Debye temperature, $T_D$, via the following formula:

$$CS(T) = IS(0) + SOD(T) \qquad (1)$$

Where *IS* stays for the isomer shift and *SOD* is the so-called second order Doppler shift i.e. a quantity related to a non-zero mean value of the square velocity of vibrations, $<v^2>$, hence a kinetic energy. Assuming the phonon spectrum obeys the Debye model, and that *IS* hardly depends on temperature, so it can be ignored [1], the temperature dependence of *CS* goes practically via the second term which is related to $T_D$ via the following relationship [2]:

$$CS(T) = IS(0) - \frac{3k_B T}{2mc}\left( \frac{3T_D}{8T} + 3\left(\frac{T}{T_D}\right)^3 \int_0^{T_D/T} \frac{x^3}{e^x - 1} dx \right) \qquad (2)$$

Here *m* stays for the mass of the Fe atom, $k_B$ is the Boltzmann constant, *c* is the speed of light, and $x = \frac{\hbar\omega}{2\pi k_B T}$ ($\omega$ being frequency of vibrations).

Determining of *CS* from a spectrum is an easy task if the spectrum is simple i.e. it is either in the form of a single line, a doublet or a sextet. It is also a relatively simple task if the spectrum is well-resolved, the number of lattice sites occupied by probe nuclei and their population on these sites are known. However, if the investigated sample has a complex crystallographic structure i.e. with several sub lattices, lower-than-cubic symmetry and, in addition, but not necessary, it is weakly magnetic thank determination of the correct value of *CS* may be quite challenging issue. Good examples of such kind of samples are so-called Frank-Kasper phases, also known as topologically close-packed (TCP) phases e. g. σ, λ, ζ, η and other [3]. Most-likely the best known case of them is sigma (σ). It has a tetragonal unit cell with 30 atoms distributed over five different lattice sites [4]. In addition, at low temperatures, it may be weakly magnetic e. g. Fe-X (X=Cr, V, Re, Mo) just to name examples of a binary Fe-based cases [5-8]. Consequently, the proper analysis of a spectrum measured on σ requires inclusion of a full Hamiltonian into the fitting procedure.

In this paper we want to introduce the simplest possible method for determining *CS*. The method is particularly well-suited for "difficult cases" like σ-phase i.e. when the number of components and their relative abundance are unknown and the spectrum has no well-resolved structure.

## 2. Simplistic method

The center shift of any Mössbauer spectrum, *CS*, is calculated using the following formula:

$$CS = \frac{\sum_{k=1}^{n}(<b> - Y^k)V^k}{\sum_{k=1}^{n}(<b> - Y^k)} \quad (3)$$

Where:

$<b> = \frac{\sum_{k=1}^{n1} Y^k + \sum_{k=n-n1+1}^{n} Y^k}{2n1}$ is the background of the spectrum, $Y^k$ stands for the number of counts in the *k*-th channel, $V^k$ is the source velocity of the *k*-th channel, *n* is the number of channels per spectrum and *n1* indicates the number of channels taken for the calculation of the <b> parameter ( the first *n1* and the last *n1 channels* out of *n* are used to calculate <b>).

Errors of <b>, Δ<b>, and of *CS*, ΔCS, are calculated based on the following formulas, respectively:

$$\Delta <b> = \left(\frac{\sum_{k=1}^{n1}(Y^k-<b>)^2 + \sum_{k=n-n1+1}^{n}(Y^k-<b>)^2}{2n1(2n1-1)}\right)^{1/2} \quad (4)$$

$$\Delta CS = CS \left(\left|\frac{\sum_{k=1}^{n}(\Delta<b>+(<b>-Y^k)^{\frac{1}{2}}))V^k}{\sum_{k=1}^{n}(<b>-Y^k)V^k}\right| + \left|\frac{\sum_{k=1}^{n}(\Delta<b>+(<b>-Y^k)^{\frac{1}{2}}))}{\sum_{k=1}^{n}(<b>-Y^k)}\right|\right) \quad (5)$$

## 3. Examples

Two examples illustrating the use of the method are given. One concerns the pyrite, $FeS_2$, which can be regarded as an "easy case", and another one depicts a σ-FeCrNi intermetallic alloy, a "difficult case" as far as the standard fitting procedure is concerned.

### 3.1. Pyrite

Unit cell of pyrite, $FeS_2$, has a cubic symmetry but Fe atoms have a slightly distorted octahedral symmetry [9], hence they experience an electric field gradient. Consequently, two hyperfine spectral parameters viz. *CS* and a quadrupole splitting, *QS,* are needed to properly analyze its Mössbauer spectrum. There is a large number of corresponding data available in the literature both for naturally occurring and synthesized pyrite. Room temperature values of *CS*, relative to that of metallic iron, spread between 0.310 mm/s and 0.329 mm/s and those of *QS* between 0.603 and 0.624 mm/s [10, 11]. The spread can be explained in terms of variations in compositions (different impurities and their content) and conditions of formation. Also differences in Mössbauer set-ups and procedures used for spectra analysis can contribute to the spread. In this study $^{57}$Fe-site spectra were recorded in a transmission mode at different temperatures (78-295 K) on a powdered sample of a natural $FeS_2$. A Co/Rh source of 14.4 keV gamma radiation was used. Figure 1 presents three spectra recorded at different temperatures. Noteworthy, the doublet-like structure persists down to 4.2 K, but a significant line broadening is observed at low temperatures related to a weak magnetism of $FeS_2$ [9].

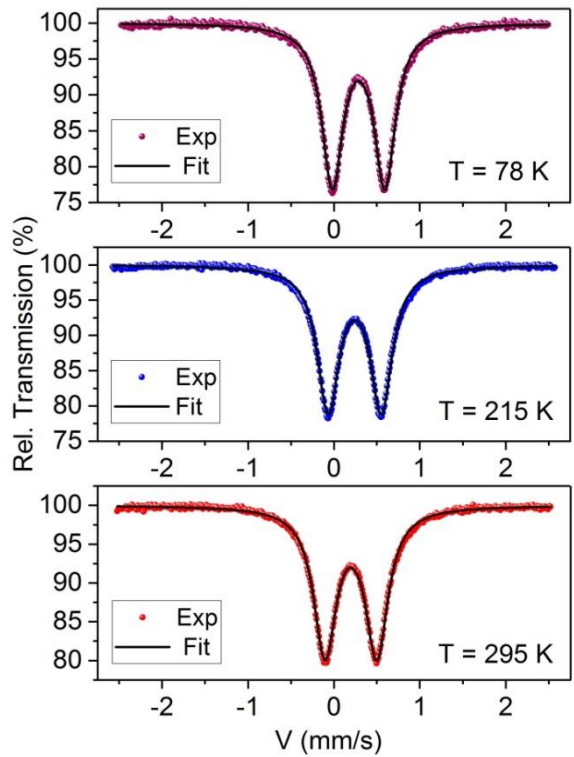

Fig. 1 Examples of the Mössbauer spectra recorded on $FeS_2$ at various temperatures. Solid lines stand for the best fit of a doublet to the data.

### 3.1.1 Analysis: Common approach

The spectra were fitted to a doublet, *CS*, *QS* and the linewidth, *G*, being free parameters. Their values at 295 K were 0.305(1) mm/s (relative to α-Fe), 0.604(1) mm/s and 0.26(1) mm/s, respectively. The temperature dependence of *CS* obtained using this fitting procedure is illustrated in Fig. 2a.

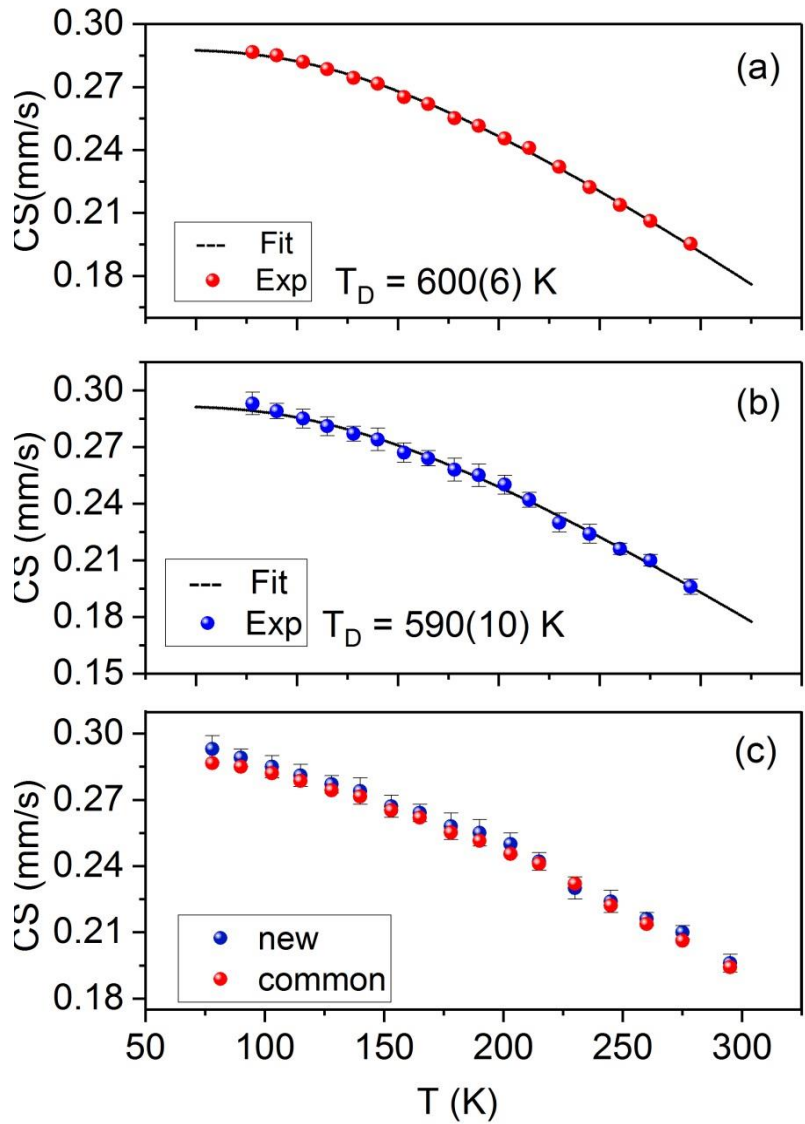

Fig. 2 Temperature dependences of the center shift, *CS*, as obtained with (a) the classic approach, (b) new approach and (c) a comparison of both. The best-fit curve to the data in terms of eq. (2) is shown as solid lines. Note: *CS*-values are given relative to a Co/Rh source.

The *CS(T)* data were fitted to eq.(2), yielding the value of 600(6) K for the Debye temperature, $T_D$.

### 3.1.2 Analysis: Simplistic approach

The values of *CS* determined by this way of spectra analysis are shown in Fig. 2b. The solid line stands for the best fit of this data to eq. (2). The obtained therefrom value of $T_D$ = 590(10) K, hence within the error limit it agrees with the one found with the classic approach. The comparison of the *CS(T)* data can be seen in Fig. 2c. One can see that a good agreement between the two sets of data exists for all temperatures. This proves that the new approach to the analysis of the spectra gives the correct values of the center shift.

It is of interest to compare the $T_D$-values obtained in this study with those obtained previously by analyzing the Mössbauer spectra. Thus, Nishihara and Ogawa reported 610(15) K obtained from the analysis of *CS(T)* measured between 77 K and 292 K [12], Kansy et. al [13] found 568(20) K for a natural sample of $FeS_2$ based on the *CS(T)*-data measured in the *T*-range between 290 K and 430 K, Polyakov et al. [11] measured a synthesized sample between 90 K and 295 K and using the *CS(T)*-data arrived at the value of 551(8.5) K. Finally, the value of $T_D$=636(5) K was determined for a natural sample by analyzing the temperature dependence of the *f*-fraction [14].

### 3.2. Sigma phase

[57]Fe-site spectra were recorded in a transmission mode at different temperatures (5-300 K) on a σ-phase $Fe_{0.525}Cr_{0.455}Ni_{0.020}$ alloy. Examples of them are shown in Fig. 3.

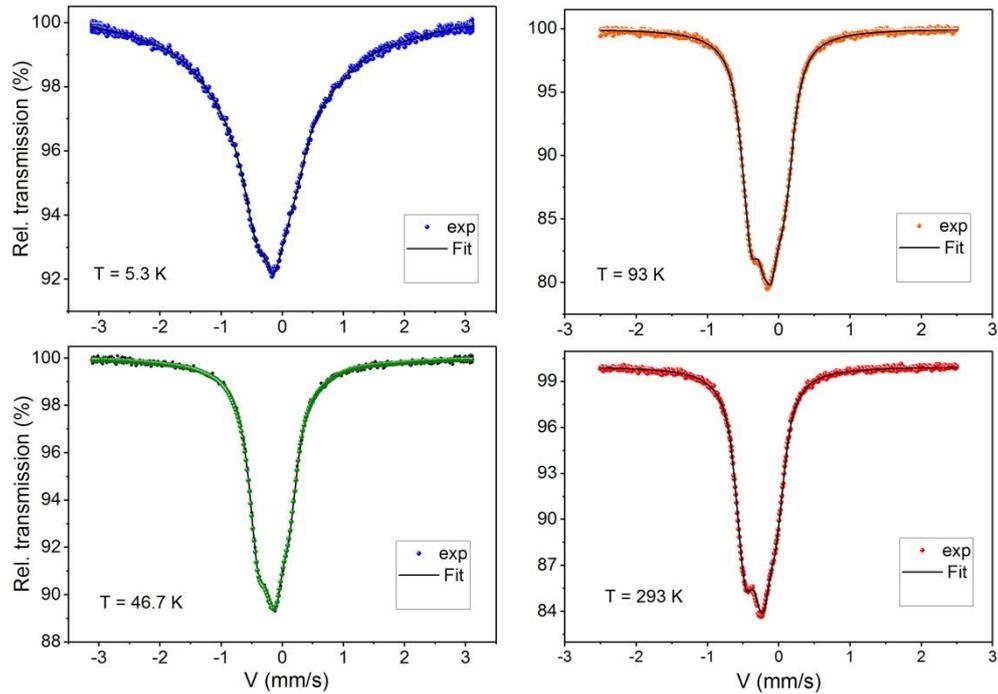

Fig. 3 Examples of the spectra recorded on the sample of σ-$Fe_{0.525}Cr_{0.455}Ni_{0.020}$ at various temperatures shown.

It is obvious that the shape of the spectra is all but not well-resolved. The spectra recorded above ~50 K are for the paramagnetic state, and for those measured below ~50 K the sample was in a magnetic phase. This means that their correct analysis of the paramagnetic case requires inclusion of the coulomb monopole interactions and the quadrupole interactions. For the magnetic case, magnetic hyperfine interactions have to be included, in addition. Furthermore, each spectrum has to be decomposed into five sub spectra as Fe atoms are present on all five lattice sites. Last, but not least, a relative contribution of the sub spectra has to be known (it can be determined by neutron diffraction (ND) experiment) because sigma-phase has no fixed stoichiometry, so the population of atoms on particular sites changes with composition. It these circumstances the correct analysis of such spectra is very challenging and requires a great skill in the analysis of the Mössbauer spectra.

### 3.2.1 Analysis: Common approach

Following the above given conditions, we analyzed the spectra in terms of five components with fixed relative contributions determined by ND. The spectra recorded in the paramagnetic phase were fitted to five doublets, whereas those measured in the magnetic phase to five Gaussians. The average values of *CS*, <*CS*>, were determined as the weighted values i.e. $<CS> = \sum_k CS_k P_k$, where *k=1-5*, $P_k$ is the relative abundance of the *k*-th component. The obtained <*CS*>(*T*)-dependence can be seen in Fig. 4a. Noteworthy, a strong anomaly is visible for T≤~50 K i.e. below the magnetic ordering temperature [15]. Consequently, the analysis of the data in terms of the eq. (2) was limited to the temperature interval 50-295 K and the value of the Debye temperature obtained therefrom is 437(7) K.

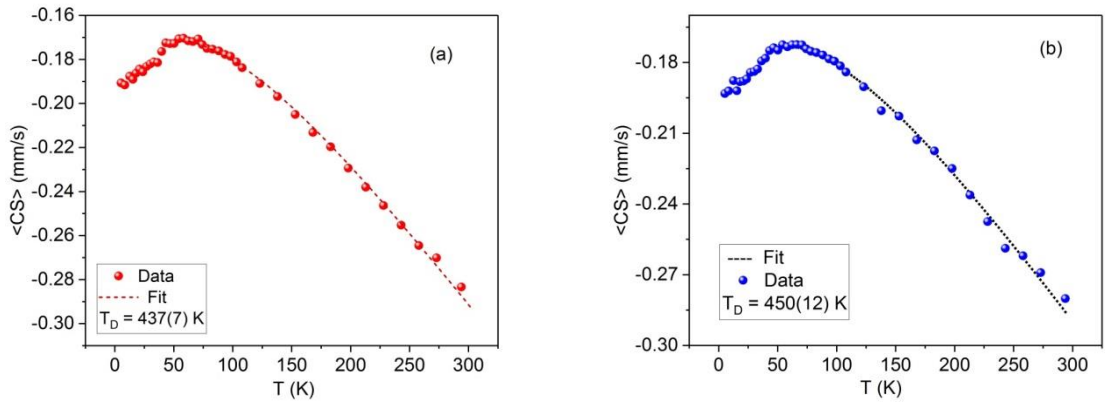

Fig. 4 Temperature dependence of the average center shift, <*CS*>, as found with (a) classic method, and (b) new method. The fit of the data to eq. (2) is marked by the dashed lines in both cases.

### 3.2.1 Analysis: Simplistic approach

The <*CS*>-data determined by this method are presented in Fig. 2b. It can be clearly seen that they are very similar to those shown in Fig. 2a. In particular, the anomaly occurring in the magnetic phase has been revealed, too. Analysis of the data in terms of eq. (2) yielded the value of 450(12) K for the Debye temperature agrees within the error limit with the corresponding value found with the classic method of analysis.

### 4. Summary

The most simple method for determining the center (isomer) shift, *CS*, of a Mössbauer spectrum is outlined. Its application has been exemplified by analysis of two sets of the spectra viz. Measured in the temperature range of 1) 78-295 K on a powder sample of a natural pyrite, $FeS_2$ and 2) 5-295 K on a powder sample of a sigma-phase $Fe_{0.525}Cr_{0.455}Ni_{0.020}$ alloy. For comparison, all the recorded spectra were also analyzed using a common procedure i.e. the spectra of $FeS_2$ were fitted to one doublet, and those of $\sigma$-$Fe_{0.525}Cr_{0.455}Ni_{0.020}$ to five sub spectra corresponding to five different lattice sites. Thus obtained sets of the *CS(T)*-data were analyzed in terms of the relevant Debye model yielding values of the Debye temperature, $T_D$. For the pyrite $T_D$=590(10) K for the new method and $T_D$=600(6) K for the common approach. The corresponding $T_D$-values found for the $\sigma$-phase sample are 437(7) K and 450(12) K, respectively. The corresponding figures are within error limit in line with each other and they give evidence that the new method, which is not only much faster, but, first off all, does not require any knowledge either on the number of sub spectra nor on their type (singlet, doublet, sextet), gives correct *CS*-values.

**Acknowledgements**

This work was financed by the Faculty of Physics and Applied Computer Science AGH UST and ACMIN AGH UST statutory tasks within subsidy of Ministry of Science and Higher Education, Warszawa.

**CRediT author statement**

**Stanisław M. Dubiel:** Conceptualization, Investigation, Validation, Formal analysis, Resources, Writing - Original Draft, Visualization: **Jan Żukrowski:** Software, Validation, Formal analysis, Investigation, Data Curation.